# Adhesive and non-adhesive contact of a rigid indenter and a thin elastic layer with surface tension

Valentin L. Popov
Technische Univetrsität Berlin, 10623 Berlin, Germany
E-mail: v.popov@tu-berlin.de

We consider an adhesive contact between a thin soft layer on a rigid substrate and a rigid cylindrical indenter ("line contact") with account of the surface tension of the layer. First, it is shown that the boundary condition for the surface outside the contact area is given by the constant contact angle – as in the case of fluids in contact with solid surfaces. In the approximation of thin layer and under usual assumptions of small indentation and small inclination angles of the surface, the problem is solved analytically.

## 1. Introduction

Classical contact mechanics as represented by the works of Hertz [1] or Bussinesq [2], see also [3], neglects the surface tension of the contacting solids. In reality, all three surfaces of bodies in contact (Fig. 1a) can be characterized by their specific surface energies $\gamma_1$, $\gamma_2$ and $\gamma_{12}$. Depending on their values, one can distinguish several cases. If the specific surface energy of the surface of elastic body outside the contact area can be neglected, then we have an adhesive contact with specific work of separation $w = \gamma_2 - \gamma_{12}$. This case was first considered in the classic work by Johnson, Kendall and Roberts [4]. If the surface energy of the elastic body outside the contact area is finite, $\gamma_1 \neq 0$, but the work of adhesion (which in the general case is equal to

$$w = \gamma_1 + \gamma_2 - \gamma_{12} \qquad (1)$$

is zero, then we have a non-adhesive contact with surface tension. One can consider such system as an elastic body coated with a stressed membrane. The corresponding theory was first developed in [5]. The general case is when both work of adhesion and surface tension of the "free surface" are finite. This leads to a general adhesive contact with surface tension. The latter attracted much interest in the last two decades in the context of indentation of soft matter (e.g. gels) [6], [7], [8]. Let us also note that another contact problem with adhesion and surface tension represents a contact of an elastic solid with a *fluid* [9]. This class differs essentially from contact of elastic bodies with surface tension and will not be discussed here.

## 2. Model

In the present paper, we consider an adhesive contact between a thin soft layer on a rigid substrate and a rigid cylindrical indenter ("line contact") with account of the surface tension of the layer Fig. 1a (left). Without consideration of the surface tension, this problem has been solved in [10]. Here we extend the study carried out in [10] to include the effect of surface tension of contacting bodies. This contact problem can be treated asymptotically exact, however, under assumptions, which can be fulfilled not for any contacting bodies. In particular, we assume that the following conditions are fulfilled: $d \ll h$, $h \ll a$, where $d$ is the indentation depth, $h$ the thickness of the layer, and $a$ the half-width of the contact (definitions see in Fig. 1a (left)). Additionally, it is assumed that the slope of the profile of contacting bodies and of the free surface outside the contact is much smaller than unity. Further conditions, if necessary, will be specified later in this paper.

Under the above conditions, the elastic layer is deformed uniaxially, independently in each point and the layer can be considered as a two-dimensional elastic foundation with effective modulus [11]

$$\tilde{E} = \frac{E(1-v)}{(1+v)(1-2v)}, \qquad (2)$$

where $E$ is elastic modulus and $v$ Poisson number. Due to local uniaxial deformation, the layer can be considered as a two-dimensional elastic foundation composed of independent springs placed with separation $\Delta x$ and $\Delta y$ correspondingly, while each spring has the stiffness

$$\Delta k = \tilde{E}\frac{A}{h}, \qquad (3)$$

with $A = \Delta x \Delta y$. When a rigid profile $f(x)$ (Fig. 1b (left)) is indented into this elastic foundation by a depth $d$, then the vertical displacements of the springs in contact are equal to

$$u_z(x) = d - f(x), \quad |x| \leq a. \qquad (4)$$

(Note that while the axis $z$ for the definition of the profile shape is directed upwards, the positive direction of the displacement $u_z(x)$ is accepted to be downwards).

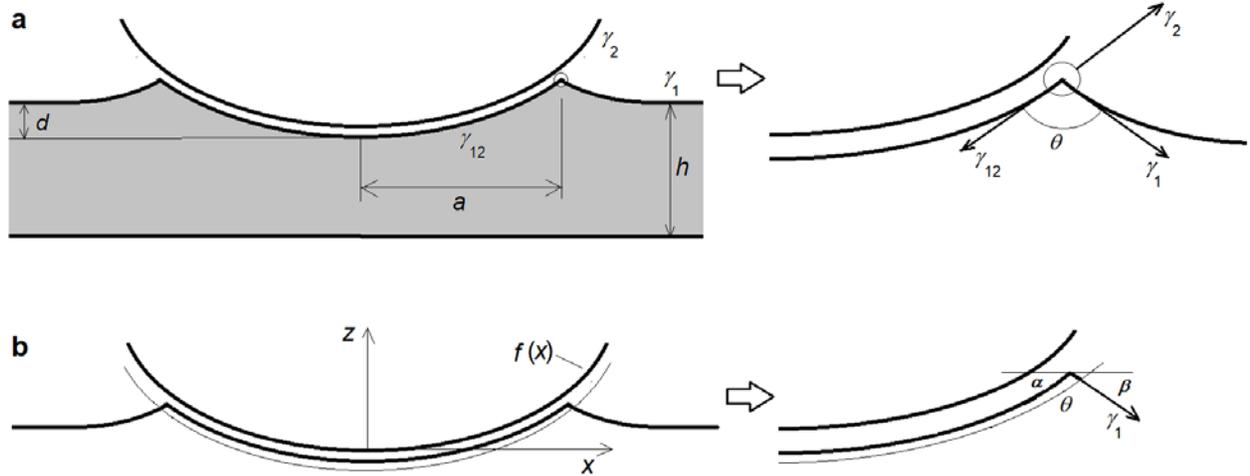

**Fig. 1 (a) left:** principal sketch of the system consisting of a rigid indenter in contact with a thin elastic layer (grey) having the initial thickness $h$. **(a) right**: Detailed view of the boundary of the contact with three surface forces corresponding to the tree surfaces meeting in the boundary. **(b) left**: Definitions of coordinates and profile of the rigid indenter as well as free body diagramm (of the system over the thin gray line). **(b) right**: Detailed picture of a part of the free body diagram showing the surfaces forces acting on the contact boundary.

The shape of the surface outside the contact area, is governed by the equation

$$\gamma_1 \frac{\partial^2 u_z(x)}{\partial x^2} = \frac{\tilde{E}}{h} u_z(x), \qquad (5)$$

which simply equates the elastic stress to the stress produced by the stressed surface (surface tension $\gamma_1$ multiplied with the surface curvature $\partial^2 u_z(x)/\partial x^2$). Solution of Eq. (5) reads

$$u_z(x) = -C \exp(-x/l), \tag{6}$$

where

$$l = \sqrt{\frac{\gamma_1 h}{\tilde{E}}}. \tag{7}$$

This length plays in the present problem the role of the "elastocapillary length".

### 3. Boundary condition at the contact boundary

Equation (5) must be completed through boundary conditions at the boundary of contact area. For deriving this boundary condition, consider a small part of the boundary encircled in Fig. 1a (right) by a gray circle. The sum of all forces acting on the boundary line parallel to the surface of the rigid body, should vanish, if the boundary friction is neglected:

$$\gamma_{12} + \gamma_1 \cos\theta - \gamma_2 = 0. \tag{8}$$

The elastic force can be neglected in this equation as it vanishes if the size of the circle tends towards zero (while the surface tensions remain constant). This equation is the same as in the case of a contact of a liquid with a solid. Karpitschka et al. come to the same conclusion by performing minimization of the complete energy functional [12].

### 4. The force acting on the rigid indenter

As we consider the line contact (no dependency on the coordinate perpendicular to the plane $(x,z)$ (not shown in Fig. 1)), it is convenient to use instead of the normal force the normal force per length, $P_N$ and relate also all other forces per unit length. The elastic force per unit length acting on the rigid indenter from the elastic layer is simply:

$$P_{el} = \frac{\tilde{E}}{h}\int_{-a}^{a} u_z(x)dx = \frac{\tilde{E}}{h}\int_{-a}^{a}(d - f(x))dx. \tag{9}$$

Apart from this elastic force, there is an additional force acting on the indenter by the surface of the elastic body outside the contact area, which is equal to (see Fig. 1b (right))

$$P_{surf} = -2\gamma_1 \sin\beta. \tag{10}$$

Under assumption that all slopes in the considered system are small, the total force acting on the rigid indenter from the elastic body is equal to

$$P_N = 2\frac{\tilde{E}}{h}\int_{0}^{a}(d - f(x))dx - 2\gamma_1 \frac{\partial u_z(x)}{\partial x}\bigg|_{x=a+0}. \tag{11}$$

In this equation, the half-width of the contact, $a$, is still not defined.

## 5. Contact half-width

The boundary condition for the surface shape can be written as

$$\left.\frac{\partial f(x)}{\partial x}\right|_{x=a-0} + \left.\frac{\partial u_z(x)}{\partial x}\right|_{x=a+0} + \theta = \pi. \quad (12)$$

Note that we assume that all slopes are small, so that the contact angle should be almost equal to $\pi$. Smaller contact angles can be realized physically but they cannot be treated in the approximation of small slopes, which is used in the present model (see, however, the next section for a more detailed discussion of the area of applicability).

At the boundary of the contact area two equations have to be fulfilled:

$$d - f(a) = -C\exp(-a/l), \quad (13)$$

and

$$\left.\frac{\partial f(x)}{\partial x}\right|_{x=a-0} + \frac{C}{l}\exp(-a/l) + \theta = \pi. \quad (14)$$

Substituting (13) as well as solution of eq. (8),

$$\pi - \theta \approx \sqrt{\frac{2w}{\gamma_1}}, \quad (15)$$

into (14) gives

$$f(a) = d + \sqrt{\frac{2wh}{\tilde{E}}} - \sqrt{\frac{\gamma_1 h}{\tilde{E}}} \cdot \left.\frac{\partial f(x)}{\partial x}\right|_{x=a-0}. \quad (16)$$

Here

$$w = \gamma_1 + \gamma_2 - \gamma_{12} \quad (17)$$

is the work of adhesion.

In the limit $\gamma_1 = 0$ (vanishing surface tension), (16) reduces to

$$f(a) = d + \sqrt{\frac{2wh}{\tilde{E}}}, \quad (18)$$

which, according to [10], is the correct result for an arbitrary profile if the surface tension is neglected.

## 6. Area of applicability of eq. (16)

Note that all above considerations are valid under assumptions listed in Section 2. In particular, the thin layer approximation can only be used if all slopes are small. For our problem, this implies that the angle $\pi - \theta$ also should be small. From Eq. (15) it then follows that the current approximation is strictly valid only if $2w \ll \gamma_1$. This means that the limit $\gamma_1 \to 0$ is not covered by the present theory. However, Eq. (18), obtained in the limit $\gamma_1 = 0$, reproduces the exact solution of the corresponding problem without surface tension. This suggests that Eq. (16) can be used as an approximate solution in the whole range of values of $0 \leq \gamma_1 \leq \infty$.

## 7. Case studies

### Case study 1: rigid plane

In this case, $f(x) = 0$ and Eq. (16) takes the form

$$0 = d + \sqrt{\frac{2wh}{\tilde{E}}}. \tag{19}$$

For negative indentation depth (which correspond to the adhesion case), this equation is fulfilled for one single value of the distance between the rigid plane and the elastic layer:

$$|d_c| = -d_c = \sqrt{\frac{2wh}{\tilde{E}}}. \tag{20}$$

If the distance becomes larger, the contact shrinks and disappears; if it becomes smaller, then it spreads to infinity. At exactly the critical value, the contact is in an indefinite equilibrium at any contact size. These properties are the same as in the case of adhesive contact without tension.

### Case study 2: flat punch with half-width $a$

As is clear from the Case study 1, a flat ended punch will detach at once at the critical distance, given by eq. (20). The surface shape outside the contact is given by

$$u_z(x) = -\sqrt{\frac{2wh}{\tilde{E}}} \exp\left(\frac{-x+a}{l}\right). \tag{21}$$

Eq. (11) for the normal force now gives

$$P_N = -2a\sqrt{\frac{2w\tilde{E}}{h}} - 2\sqrt{2w\gamma_1} = -2^{3/2}w^{1/2}\left(a\sqrt{\frac{\tilde{E}}{h}} + \sqrt{\gamma_1}\right). \tag{22}$$

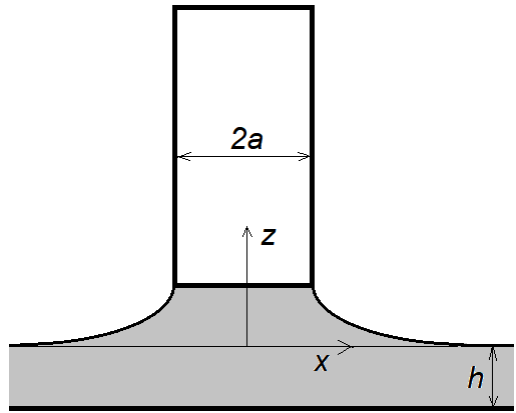

**Fig. 2** Adhesive contact with surface tension of a flat rigid indenter with a thin elastic layer.

The force of adhesion is minus the normal force acting on the rigid indenter:

$$P_A = 2^{3/2} w^{1/2} \left( a \sqrt{\frac{\tilde{E}}{h}} + \sqrt{\gamma_1} \right). \tag{23}$$

We see that, at the given work of separation $w$, the surface tension leads to an increase of the force of adhesion. The contact configuration is illustrated in Fig. 2.

*Case study 3: wedge shape*

If the shape of the rigid indenter is given by $f(x) = |x| \tan \delta$, then eq. (16) takes the form

$$a \tan \delta = d + \sqrt{\frac{2wh}{\tilde{E}}} - \sqrt{\frac{\gamma_1 h}{\tilde{E}}} \cdot \tan \delta \tag{24}$$

and the half-width of the contact area is given by

$$a = (\tan \delta)^{-1} \left( d + \sqrt{\frac{2wh}{\tilde{E}}} \right) - \sqrt{\frac{\gamma_1 h}{\tilde{E}}}. \tag{25}$$

The shape of the surface outside the contact is given by

$$u_z(x) = \left( -\sqrt{\frac{2wh}{\tilde{E}}} + \sqrt{\frac{\gamma_1 h}{\tilde{E}}} \cdot \tan \delta \right) \exp\left( \frac{-x+a}{l} \right). \tag{26}$$

For the normal force (Eq. (11)) we get:

$$P_N = 2 \frac{\tilde{E}}{h} \left( ad - \frac{a^2}{2} \tan \delta \right) + 2 \left( -\sqrt{2w\gamma_1} + \gamma_1 \cdot \tan \delta \right) \tag{27}$$

Solving Eq. (24) with respect to $d$:

$$d = a \tan \delta - \sqrt{\frac{2wh}{\tilde{E}}} + \sqrt{\frac{\gamma_1 h}{\tilde{E}}} \cdot \tan \delta \tag{28}$$

and inserting this result into (27) gives the force per length as function of the contact half-width $a$:

$$P_N = 2 \frac{\tilde{E}}{h} \left( \frac{a^2}{2} \tan \delta - a \left( \sqrt{\frac{2wh}{\tilde{E}}} - \sqrt{\frac{\gamma_1 h}{\tilde{E}}} \cdot \tan \delta \right) \right) + 2 \left( -\sqrt{2w\gamma_1} + \gamma_1 \cdot \tan \delta \right). \tag{29}$$

Minimizing with respect to $a$, gives the adhesion force

$$F_A = |P_{N,\min}| = \frac{2w}{\tan \delta} - \gamma_1 \tan \delta. \tag{30}$$

For small $\gamma_1$, the surface tension leads to a decrease of the force of adhesion.

*Case study 4: parabolic shape*

Let us consider the special case of a parabolic profile

$$f(x) = \frac{x^2}{2R}. \tag{31}$$

Equation (16) now takes the form

$$\frac{a^2}{2R} = d + \sqrt{\frac{2wh}{\tilde{E}}} - \sqrt{\frac{\gamma_1 h}{\tilde{E}}} \frac{a}{R} \tag{32}$$

Its solution with respect to $a$ reads

$$a = -\sqrt{\frac{\gamma_1 h}{\tilde{E}}} + \sqrt{\frac{\gamma_1 h}{\tilde{E}} + 2R\left(d + \sqrt{\frac{2wh}{\tilde{E}}}\right)}. \tag{33}$$

In the case of vanishing surface tension, $\gamma_1 = 0$, eq. (33) reduces to $a = \sqrt{2R\left(d + \sqrt{\frac{2wh}{\tilde{E}}}\right)}$ meaning that the contact boundary is defined by cutting the profile at the height $d + \sqrt{\frac{2wh}{\tilde{E}}}$ which coincides with the result of paper [10] for the corresponding problem with vanishing surface tension.

## 8. Non-adhesive contact

Let us consider the limiting case of non-adhesive contact with tension separately. Under "no-adhesive" contact we will understand the contact of surfaces with vanishing work of separation, $w = 0$. From the definition (1), it follows that in this case

$$\gamma_1 = \gamma_{12} - \gamma_2. \tag{34}$$

From (8), it the follows that

$$\cos\theta = \frac{\gamma_2 - \gamma_{12}}{\gamma_1} = -1 \tag{35}$$

and $\theta = \pi$. This means that the slope is continuous at the boundary of the contact. Eq. (14) takes now the form

$$\left.\frac{\partial f(x)}{\partial x}\right|_{x=a-0} + \frac{C}{l}\exp(-a/l) = 0. \tag{36}$$

Taking (13) into account, we come to the equation

$$d - f(a) = l \cdot \left.\frac{\partial f(x)}{\partial x}\right|_{x=a-0} = \sqrt{\frac{\gamma_1 h}{\tilde{E}}} \cdot \left.\frac{\partial f(x)}{\partial x}\right|_{x=a-0}. \tag{37}$$

Consider as example a contact of a parabolic indenter $f(x) = x^2/(2R)$. Eq. (37) takes the form

$$a^2 + 2al - 2Rd = 0. \tag{38}$$

For the contact radius we thus get

$$a = -l + \sqrt{2Rd + l^2}\ . \tag{39}$$

This means that the surface tension leads to a decrease of the contact width compared with the non-adhesive contact without surface tension.

For the total normal force we get according to (11)

$$P_N = \frac{2\tilde{E}}{3Rh}\left[\left(2Rd + l^2\right)^{3/2} - l^3\right] = \frac{2\tilde{E}}{3Rh}\left[\left(2Rd + \frac{\gamma_1 h}{\tilde{E}}\right)^{3/2} - \left(\frac{\gamma_1 h}{\tilde{E}}\right)^{3/2}\right]. \tag{40}$$

For given $d$, the normal force with surface tension is larger than that without surface tension.

## 9. Conclusion

In the present paper, we considered a general adhesive contact of a thin elastic body with a rigid indenter. An important conclusion is that at the boundary of the contact area, the surface of the elastic layer meets the surface of the rigid indenter under a fixed contact angle, which is determined uniquely by the specific surface energies of the rigid body, the elastic body and the interface. The solution obtained for very small surface tension (compared with the work of adhesion) seems to provide a good approximation for arbitrary values of the surface tension. In the case of a non-adhesive contact, surface tension makes the contact stiffer (at the given indentation depth, the contact half-width becomes smaller and the indentation force larger). In the case of adhesive contact, the influence of surface tension seems to be more complicated: In the case of a flat-ended punch, it increases with increasing the surface tension, while in the case of a wedge, it decreases. Thus, the influence of the surface tension on the adhesion force seems to be dependent on the particular geometry of the contacting bodies.

## 10. Acknowledgements

The author is thankful to Qiang Li, Weike Yuan and Iakov Lyashenko for helpful discussions. This work was financially supported by the DFG (project number PO 810/55-1).